\documentstyle[12pt]{article}
\makeatletter

\ifcase \@ptsize
\or 
\or 
\fi
\advance\textheight by \topskip

\ifcase \@ptsize
\or 
\or 
\fi

\def\section{\@startsection {section}{1}{\z@}{-3.5ex plus -1ex minus 
 -.2ex}{2.3ex plus .2ex}{\large\bf\centering}}
\def\subsection{\@startsection{subsection}{2}{\z@}{-3.25ex plus -1ex minus 
 -.2ex}{1.5ex plus .2ex}{\sc}}
\def\@cite#1#2{\nolinebreak$^{[\scriptstyle #1\if@tempswa , #2\fi]}$}
\def\@citex[#1]#2{\if@filesw\immediate\write\@auxout{\string\citation{#2}}\fi
  \def\@citea{}\@cite{\@for\@citeb:=#2\do
    {\@citea\def\@citea{,\penalty\@m}\@ifundefined
       {b@\@citeb}{{\bf ?}\@warning
       {Citation `\@citeb' on page \thepage \space undefined}}%
{\csname b@\@citeb\endcsname}}}{#1}}
\@definecounter{footnote}

\gdef\@publabel{\hfil}
\gdef\@pubdate{\null}
\gdef\@pubnumber{\null}
\gdef\@author{\null}
\gdef\@title{\null}
\gdef\@abstract{\null}
\long\def\pubdate#1{\gdef\@pubdate{#1}}
\long\def\pubnumber#1{\gdef\@pubnumber{#1}}
\long\def\publabel#1{\gdef\@publabel{#1}}
\long\def\author#1{\gdef\@author{#1}}
\long\def\title#1{\gdef\@title{#1}}
\long\def\abstract#1{\gdef\@abstract{#1}}
\def\titlerelax{
\let\maketitle\relax
\let\settitleparameters\relax
\let\consolidatetitle\relax
\let\inittitlepage\relax
\let\finishtitlepage\relax
\let\titlepagecontents\relax
\let\multithanks\relax
\let\titlebaselines\relax
\let\@makepub\relax
\let\@maketitle\relax
\let\@makeauthor\relax
\let\@makeabstract\relax
\let\@maketitlenote\relax
\let\thanks\relax
\let\titlerelax\relax}
\def\titleclean
{\gdef\@titlenote{}
\gdef\@abstract{}
\gdef\@author{}
\gdef\@title{}
\gdef\@pubdate{}\gdef\@pubnumber{}\gdef\@publabel{}
\gdef\@dpublabel{}
}

\def\@makepub{\vbox to \z@{\hbox to \textwidth{\hfill
\@publabel \hfill
\llap{\parbox[t]{0.25\textwidth}{\raggedleft\@pubnumber}}}%
\vss}}
\def\@maketitle{\vskip 60pt \begin{center}
 {\LARGE \@title \par}
 \end{center}}
\def\@makeauthor{{\def\and{\smallskip {\normalsize \rm and\smallskip}}
\long\def\address##1{{\def\and{\\and\\}\medskip
				{\small \it \\##1\\}
}}
{\centering
 \vskip 2.5em
 \large \lineskip .75em
 \@author}
 \par}} 
\def\@makedate{\vskip 1.5em 
 {\raggedright \small \noindent\@pubdate \par}}
\def\@makeabstract{\vskip 1.5em
{\small 
\begin{center}
{\bf ABSTRACT\vspace{-.5em}\vspace{0pt}} 
\end{center}
\quotation \@abstract \endquotation}}
\def\maketitle{
\let\footnotesize\small \setcounter{page}{1}
\@makepub
\@maketitle
\@makeauthor
\@makeabstract
\@thanks
\@makedate
\setcounter{footnote}{0}
\vspace{-40pt}
}

\begin{document}
\newcommand{\hsp}{\hspace{0.08in}}
\newcommand{\be}{\begin{equation}}
\newcommand{\ee}{\end{equation}}
\newcommand{\beaa}{\begin{eqnarray*}}
\newcommand{\eeaa}{\end{eqnarray*}}
\newcommand{\shs}{\shortstack}
\newcommand{\scs}{\scriptsize}
\newcommand{\fns}{\footnotesize}
\newcommand{\la}{\lambda}
\newcommand{\om}{\omega}

\bibliographystyle{npb}

\pubnumber{hep-th/9608055 \ DAMTP-96-70 \\\today}

\title{Remarks on excited states of affine Toda solitons}
\author{G.M. Gandenberger\footnote{{\tt G.M.Gandenberger@damtp.cam.ac.uk}.
\linebreak
Address from 1st October: Dept of Physics, Brandeis Univ., Waltham, MA
02254-9110, USA} 
\linebreak
N. J. MacKay\footnote{{\tt N.J.Mackay@damtp.cam.ac.uk}.
Stokes Fellow, Pembroke College}
\address{Dept of Applied Maths and Theoretical Physics, \linebreak
         Cambridge University, \linebreak
         Cambridge, CB3 9EW, \linebreak
         England}}
\abstract{
The identification in affine Toda field theory of the quantum
particle with the lowest breather allows us to re-interpret
discrete modes of excitation of solitons as breathers bound
to solitons, and thus to investigate them through the proposed
soliton-breather S-matrices. There are implications for the 
physical spectrum and for the semiclassical soliton mass corrections.
}

\maketitle
\baselineskip 18pt
\parskip 12pt
\parindent 10pt

\section{Introduction}

In a recent series of papers\cite{gande95,gande95b,gande95c} 
(to which the reader is referred for background and notation)
we have constructed S-matrices invariant under a variety
of quantized affine algebras, and have conjectured that
they describe quantum affine Toda solitons\cite{hollo92,olive93} 
and their bound states.
This is a strong conjecture to make, because the Hamiltonian
of the theory is complex, and although the solitons have real
energies, momenta and higher-spin conserved charges, the quantum
theory would appear non-unitary. Further, classically the solitons 
may be unstable\cite{khast95}, and have topological charges
which do not fill the relevant multiplets. It is quite possible
that only in some suitably restricted sector can the classical
solitons be quantized, or our exact $S$-matrices describe them.
Nevertheless there is much commonality of structure, and 
we proceed on the assumption that the spectrum of affine
Toda theory (with its solitons, excited solitons, breathers and 
particles) is what the S-matrices describe, which leads in each case
to the identification of the lowest scalar soliton bound state or 
`breather' with the quantum particle. In this letter we examine the 
implications for the semiclassical discussion of the solitons.

In the semiclassical calculation of the soliton 
masses\cite{hollo93b,deliu94b,macka94},
the zero-point energies of all the modes of vibration
of the solitons were summed. These included a set of discrete
modes, which could be regarded as a particle bound to a soliton.
Some of these discrete modes are linearizations of the excited 
solitons present in the spectrum. Should the rest also be identified
with physical states, to be invited to the bootstrap? If the particle
is identical to the lowest breather, we can attempt to provide an answer  
through the pole structure of the soliton-breather S-matrices.
We describe this in the next section.

The semiclassical mass calculations had unresolved problems
for certain solitons for nonsimply-laced untwisted algebras.
Specifically, there was a multiplicity in the zero of the
transmission coefficient of a wave incident on a soliton
and thus in the frequency of the associated discrete mode.
Had these been counted with multiplicity, the masses would then have 
agreed with the soliton S-matrices; yet despite several attempts
only one such mode for each eigenvalue could be explicitly constructed.
In the soliton-breather S-matrices, however, the frequencies
of the discrete modes can be read off exactly, and we find that
the degeneracies  are split by higher-order contributions in
the coupling constant, so that the full multiplicity of modes must
therefore exist. We describe this in the last section.

\section{Discrete modes and S-matrix poles}

Suppose there is a discrete mode solution to the equation
of motion for a small perturbation to a soliton of type $a$,
and that this solution takes the asymptotic form of a particle 
of type $b$ at $x\rightarrow -\infty$, and has frequency
$\nu = m_b \sin\phi$.
Now let us identify the particle with the lowest breather state
$B^{(b)}_1$ and denote the soliton of type $a$ as $A^{(a)}$.
If $S_{B^{(b)}_1A^{(a)}}$ has a pole at rapidity difference 
$\theta={i\pi\over 2} -i\phi+{\cal O}(\beta^2)$ (where $\beta$ is 
the coupling constant) then the mass of the intermediate state
is $m_b \sin\phi+{\cal O}(\beta^2)$ more than the mass of the 
original soliton, and we might expect the discrete modes to 
coincide with such poles.

\pagebreak
For $d_{n+1}^{(2)}$ this is indeed what we find.
The possibility of constructing a classical bound
state is signalled by a zero in the transmission 
coefficient for a wave scattering off a soliton, and 
comparing the expressions for the transmission
coefficients\cite{fring94,macka94} (which are the phase factors
$X_{a\,b}(\eta)$ for solitons of rapidity difference $\eta$)
with those for the 
S-matrices\cite{gande95b}, we find that for each zero/pole
in $X_{a\,b}(i\phi)$, there is precisely one pole/zero 
(plus its cross-channel partner) in
$S_{B^{(b)}_1A^{(a)}}({i\pi\over 2} -i\phi+{\cal
O}(\beta^2))$.\footnote{This is in contrast to soliton-soliton 
scattering, where, at least for $d_3^{(2)}$, the correspondence
is between the zeros/poles in $X_{a\,b}(i\phi)$ and poles/zeros
in $S_{A^{(a)}A^{(b)}}(i\phi)$.}

Thus all the discrete modes
(including those putative bound states whose existence was 
forbidden classically by their bad asymptotic behaviour at 
$x\rightarrow \infty$) are contained
among, though generally they by no means exhaust, 
the relevant ({\em i.e.\ }$0<\phi< {\pi\over 2}$) simple poles in the
S-matrix. Among the poles associated with the allowed classical
bound states, in turn, are found those already included in the
bootstrap as excited solitons. (Recall that a pole's being simple 
is not alone enough for its
inclusion as a physical state: a generalized Coleman-Thun mechanism
can lead to on-shell higher-order diagrams which nevertheless
correspond to simple poles in the S-matrix.)

Consider, as an example, the $d^{(2)}_3$
theory\cite{gande95b}\footnote{A more detailed discussion of the
$d_3^{(2)}$ pole structure can be found in \cite{gandethesis}.}.
There are two species. The transmission coefficients are
\beaa
X_{1\,1}(\eta) & = &  \{1\}\{-1/2\}  \\
X_{1\,2}(\eta)=X_{2\,1}(\eta) & = & \{\sqrt{3}/2\}  \\
X_{2\,2}(\eta) & = & \{1\}\{1/2\}
\eeaa
where 
$$
\{a\} = {\cosh\eta - a \over \cosh\eta + a} \;,
$$
and the condition for a zero in $X$ to give a normalizable
bound state is that $m_b\cos\phi<m_a$. Thus there are two
allowed discrete modes, for $a=2$ and $b=1,2$, whilst that
for $a=1$ and $b=2$ is disallowed.

The necessary S-matrices are
\beaa
S_{B^{(1)}_1A^{(1)}}(\theta) & = & 
\left( -{\om \over 2}\right)\left(-{5\om \over 2}-1\right)
\left( {3\om \over 2}\right)\left( {3\om \over 2}+1\right)
\\[0.2in]
S_{B^{(1)}_1A^{(2)}}(\theta) = S_{B^{(2)}_1A^{(1)}}(\theta) & = & 
(2\omega)(\omega+1) 
\\[0.1in]
S_{B^{(2)}_1A^{(2)}}(\theta) & = &
\left( {5\omega\over 2}\right)\left( {\omega\over 2}+1\right)
\left( {3\omega\over 2}\right)\left( {3\omega\over 2}+1\right) \;,
\eeaa
in which we have used the notation
$$
(y) \equiv {\sin\left({\pi\over h\la}(\mu+y)\right) \over 
\sin\left({\pi\over h\la}(\mu - y)\right) }\;, \hspace{0.3in}
\la\equiv{4\pi\over\beta^2}-{4\over 3}\;, \hspace{0.3in}
\om\equiv {2\pi\over\beta^2}-1 \;, \hspace{0.3in} \mu \equiv
-i{h\la \over 2\pi}\theta \hspace{0.2in}{\rm and}
\hspace{0.2in}h=3\;, 
$$
and we see that for each $\{\sin(a\pi/6)\}$ in $X$
we find one factor $(a\omega/2+{\cal O}(1))$, and 
its cross channel partner, in $S$.

Of the relevant poles, first note that those at 
$\mu = {3\om \over 2}+1$, in $S_{B^{(1)}_1A^{(1)}}$ and 
$S_{B^{(2)}_1A^{(2)}}$, correspond to the fusion into elementary 
solitons $A^{(1)}$ and $A^{(2)}$ respectively, whilst
the poles at $\mu = {3\om \over 2}$ are their cross channel partners.
Classically they correspond to zero modes, at $\phi=0$. 
Of the poles at $\mu = \om + 1$, that in
$S_{B^{(1)}_1A^{(2)}}$ corresponds to the excited soliton
$A^{(2)}_1$, whilst that in $S_{B^{(2)}_1A^{(1)}}$ is
classically the discrete mode disallowed by its bad asymptotic
behaviour. In the quantum theory this pole was not included in the
bootstrap but could be explained by the following
higher(-loop)-order diagram, in which the S-matrix denoted by the
black dot has a simple zero:
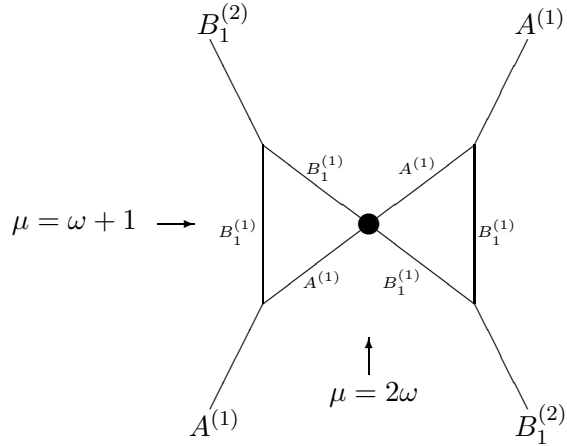
\begin{center}
\begin{picture}(130,160)
\put(0,0){\line(1,2){20}}
\put(120,0){\line(-1,2){20}}
\put(20,40){\line(0,1){60}}
\put(20,40){\line(4,3){80}}
\put(20,100){\line(4,-3){80}}
\put(100,40){\line(0,1){60}}
\put(20,100){\line(-1,2){20}}
\put(100,100){\line(1,2){20}}
\put(60,70){\circle*{8}}
\put(-8,-10){\shs{\fns{$A^{(1)}$}}}
\put(115,-10){\shs{\fns{$B^{(2)}_1$}}}
\put(115,142){\shs{\fns{$A^{(1)}$}}}
\put(-5,142){\shs{\fns{$B^{(2)}_1$}}}
\put(3,65){\shs{\tiny{$B^{(1)}_1$}}}
\put(101,65){\shs{\tiny{$B^{(1)}_1$}}}
\put(35,46){\shs{\tiny{$A^{(1)}$}}}
\put(65,46){\shs{\tiny{$B^{(1)}_1$}}}
\put(36,89){\shs{\tiny{$B^{(1)}_1$}}}
\put(71,89){\shs{\tiny{$A^{(1)}$}}}
\put(60,13){\vector(0,1){14}}
\put(45,3){\shs{\fns{$\mu = 2\om$}}}
\put(-20,70){\vector(1,0){14}}
\put(-75,68){\shs{\fns{$\mu = \om+1$}}}
\end{picture}
\end{center}
The interesting pole is that at $\mu = {\omega\over 2}+1$ in
$S_{B^{(2)}_1A^{(2)}}$, since this corresponds to an allowed
classical bound state and yet was not included in the bootstrap
as a physical state, but was explained by the higher-order
diagram shown below. (The pole at $\mu = {5\om \over 2}$
corresponds to the same process in the cross channel, as indicated by
the arrows in the diagram.)
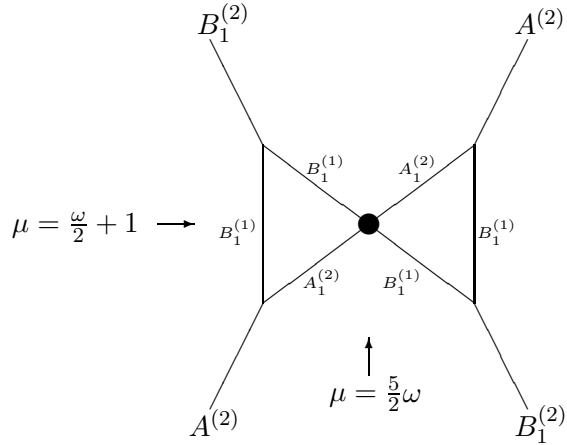
\begin{center}
\begin{picture}(130,160)
\put(0,0){\line(1,2){20}}
\put(120,0){\line(-1,2){20}}
\put(20,40){\line(0,1){60}}
\put(20,40){\line(4,3){80}}
\put(20,100){\line(4,-3){80}}
\put(100,40){\line(0,1){60}}
\put(20,100){\line(-1,2){20}}
\put(100,100){\line(1,2){20}}
\put(60,70){\circle*{8}}
\put(-8,-10){\shs{\fns{$A^{(2)}$}}}
\put(115,-10){\shs{\fns{$B^{(2)}_1$}}}
\put(115,142){\shs{\fns{$A^{(2)}$}}}
\put(-5,142){\shs{\fns{$B^{(2)}_1$}}}
\put(3,65){\shs{\tiny{$B^{(1)}_1$}}}
\put(101,65){\shs{\tiny{$B^{(1)}_1$}}}
\put(35,46){\shs{\tiny{$A^{(2)}_1$}}}
\put(65,46){\shs{\tiny{$B^{(1)}_1$}}}
\put(36,89){\shs{\tiny{$B^{(1)}_1$}}}
\put(71,89){\shs{\tiny{$A^{(2)}_1$}}}
\put(60,13){\vector(0,1){14}}
\put(45,3){\shs{\fns{$\mu = \frac 52\om$}}}
\put(-20,70){\vector(1,0){14}}
\put(-75,67){\shs{\fns{$\mu = \frac{\om}2 +1$}}}
\end{picture}
\end{center}
Of course, this does not forbid the pole's inclusion in the bootstrap;
it merely shows it not to be necessary. In the absence of a way
of summing diagrams and relating them to S-matrix residues, as is
possible for the particle S-matrices\cite{brade90}, we cannot prove
that these poles should not be invited to the bootstrap.
Nevertheless we believe that they are not, and that the spectrum
given previously\cite{gande95b} is complete. Poles in our S-matrices
seem to correspond to precisely one on-shell diagram: for example,
in those poles which are invited to the bootstrap we do not
find further higher-order diagrams to confuse the issue.

In the above diagram we note that all the internal states are
expected to be quantized forms of classical solutions. The classical 
bound state might therefore be expected to be the linearized
form of the classical solution at any time during the process ({\em
i.e.\ }corresponding to any vertical slice
through the diagram). Presumably the same then applies to
the pole considered on the previous page, 
at $\mu = \om + 1$ in $S_{B^{(2)}_1A^{(1)}}$, for which the 
classical mode is disallowed. It is not clear
what distinguishes the two at S-matrix level.

\section{Lifting of degeneracy in discrete modes}

The theories based on untwisted nonsimply-laced algebras are
more complicated in several ways. The construction of S-matrices was
only possible relatively recently, with the advent of solutions of
the Yang-Baxter equation for twisted algebras\cite{deliu95b,gande95c},
and there are outstanding problems with the semiclassical soliton
masses.

In such theories the solitons are obtained by `folding' the Dynkin
diagram of a (`parent') simply-laced theory. Most of the solitons
have single parents, but some are obtained by folding parent
multisolitons, and (solely) in the case of particles traversing
such multisolitons, the zero in $X$ occurs with multiplicity.
However, the corresponding
multiplicity of classical solutions could not be found, so the
contributions were counted only once. As pointed out in the notes
to the original papers\cite{macka94,deliu94b}, the full
multiplicity is needed for agreement with the soliton mass ratios
predicted by the exact S-matrices, and somehow the degeneracy has
to be broken. 

Using the breather-particle correspondence, we can check 
whether this happens, since the S-matrix poles are exact to all
orders in $\beta^2$. Let us take as our example the $b^{(1)}_n$
theory, where the $n$th soliton is folded from the $n$th and $n\!+\!1$th
solitons in the $d_{n+1}^{(1)}$ theory. The transmission 
coefficient for particle $a$ through soliton $n$ is $\{\sin(a\pi/2n)\}^2$,
and the mode with frequency $m_a\cos\left({a\pi\over 2n}\right)$
thus appears with a problematic double zero.

The relevant $S$-matrix is
\beaa
S_{B^{(a)}_1A^{(n)}} & = &
\prod_{k=1}^a 
\left( {\omega\over 2}(2k-a) + {3\over 4}\right)
\left( {\omega\over 2}(2n-2k+a) - {1\over 4}\right) \\
& & \hspace{0.5in}\times
\left( {\omega\over 2}(2-2k+a) + {1\over 4}\right)
\left( {\omega\over 2}(2n-2+2k-a) + {1\over 4}\right) \;,
\eeaa
where now
$$
\la\equiv{4\pi\over\beta^2} -{2n-1\over 2n}\;, \hspace{0.3in} \omega\equiv
{4\pi\over\beta^2}-1 \;, \hspace{0.3in} \mu = -i{h\la \over 2\pi}\theta
\hspace{0.2in} {\rm and}\hspace{0.2in} h=2n\;,
$$
and we wish to examine
the poles at $\theta={ai\pi\over 2n}+{\cal O}(\beta^2)$. What we
find is not a double pole but two closely separated simple poles,
at $\mu = {a\over 2}\om+{1\over 4}$ and $\mu = {a\over
2}\om+{3\over 4}$ (which for general $a$ we expect to correspond
to higher-order diagrams). 
Thus we expect that the frequencies, degenerate at ${\cal O}(1)$, are 
separated at next order and should therefore be counted with multiplicity.

This does not settle the question of the validity of the semiclassical
approximation; it merely shows the S-matrix and our interpretation
of it to be self-consistent. We still do not know whether the
correct multiplicity of distinct solutions can be constructed
semiclassically\footnote{Some effort has failed to produce
them\cite{macka96}, but not yet enough for us to suggest that they 
do not exist. In particular, our boundary conditions may have
been too constraining, with the problematic cases being especially 
sensitive to the regularizing of the system by its being placed in a 
box.}, or whether instead some demisemiclassical method is required, 
producing solutions indistinguishable at semiclassical order. Nor, if 
some such method were to be used, could we yet characterize from 
S-matrix information alone which poles correspond to discrete modes,
or whether they have allowed asymptotic behaviour.

\vspace{0.4in}
{\bf Acknowledgments}

NJM would like to thank Robert Weston and G\'erard Watts for useful 
conversations, and G\'erard Watts for a critical reading of the manuscript.
GMG acknowledges financial support from the Cambridge Kurt Hahn Trust.


\vspace{0.4in}
\baselineskip 16pt

\begin{thebibliography}{10}
%
\bibitem{gande95} 
G.M. Gandenberger, 
{\em Exact S-matrices for bound
states of $a_2^{(1)}$ affine Toda solitons}, 
\newblock Nucl.~Phys.\ {\bf B449} (1995), 375,
{\tt hep-th/9501136}
%
\bibitem{gande95b}
G.M.~Gandenberger and N.J.~MacKay,
\newblock {\em Exact $S$-matrices for $d^{(2)}_{n+1}$ affine Toda
solitons and their bound states},
\newblock Nucl.~Phys.\ {\bf B457} (1995), 240, {\tt hep-th/9506169}
%
\bibitem{gande95c}
G.M.~Gandenberger, N.J.~MacKay and G.M.T.~Watts,
\newblock {\em Twisted algebra $R$-matrices and exact $S$-matrices for
$b^{(1)}_{n}$ affine Toda
solitons and their bound states},
\newblock Nucl.~Phys.\ {\bf B465} (1996), 329, {\tt hep-th/9509007}
%
\bibitem{hollo92}
T.J. Hollowood, 
{\em Solitons in affine Toda field
theory}, Nucl. Phys. {\bf B384} (1992), 523;
{\em Quantizing $sl(N)$ solitons and
the Hecke algebra}, Int.Jour.Mod.Phys. {\bf A8} No.5 (1993), 947,
{\tt hep-th/9203076}
%
\bibitem{olive93} D.I. Olive, N. Turok and J.W.R. Underwood, 
{\em Solitons
and the energy-momentum tensor for affine Toda theory}, Nucl.Phys.
{\bf B401} (1993), 663; and {\em Affine Toda solitons and vertex
operators}, Nucl. Phys. {\bf B409} (1993) [FS], 509, {\tt hep-th/9305160};

M. Freeman, {\em Conserved charges and soliton
solutions in affine Toda theory},
Nucl. Phys. {\bf B433} (1995), 657, {\tt hep-th/9408092}
%
\bibitem{khast95}
S. Pratik Khastgir and R. Sasaki,
{\em Instability of solitons in imaginary-coupling affine Toda field
theory}, Prog.~Theor.~Phys.\ {\bf 95} (1996),485, {\tt hep-th/9507001} 
%
\bibitem{hollo93b} 
T.J. Hollowood, 
{\em Quantum soliton mass
corrections in $sl(n)$ affine Toda field
theory}, Phys.~Lett.\ {\bf B300} (1993), 73, {\tt hep-th/9209024}
%
\bibitem{deliu94b} 
G.W. Delius and M.T. Grisaru, {\em Toda soliton mass
corrections and the particle-soliton duality conjecture},
Nucl.~Phys.\ {\bf B441} (1995), 259, {\tt hep-th/9411176}
\bibitem{macka94}
 N.J. MacKay and G.M.T. Watts, {\em Quantum mass
corrections for affine Toda solitons},
Nucl.~Phys.\ {\bf B441} (1995), 277, {\tt hep-th/9411169};

G.M.T. Watts, {\em Quantum mass corrections for $c_2^{(1)}$ affine
Toda solitons}, Phys.~Lett.\ {\bf B338} (1994), 40, {\tt hep-th/9404065}
%
\bibitem{fring94} 
A.Fring, P.R. Johnson, M.A.C.Kneipp and D.I. Olive,
{\em Vertex operators and soliton time delays in affine Toda field theory},
Nucl.~Phys.\ {\bf B430} (1994), 597, {\tt hep-th/9405034}
%
\bibitem{gandethesis}
G.M.~Gandenberger, {\em Exact $S$-matrices for quantum affine Toda
solitons and their bound states}, Ph.D.\ thesis, Cambridge University
1996, {\em unpublished}  
%
\bibitem{brade90} 
H.W. Braden, E. Corrigan, P.E. Dorey and R. Sasaki,
{\em Affine Toda field theory and exact S-matrices}, Nucl.~Phys.\ {\bf
B338} (1990), 689; and {\em Multiple poles and other features of
affine Toda field theory}, Nucl.~Phys.\ {\bf B356} (1991), 469
%
\bibitem{deliu95b}
G.W.~Delius, M.D.~Gould and Y.-Z.~Zhang,
\newblock {\em Twisted quantum affine algebras and solutions to the
Yang--Baxter equation},
Int.~J.~Mod.~Phys.\ {\bf A11} (1996), 3415, {\tt q-alg/9508012}
%
\bibitem{macka96}
N.J. MacKay and G.M.T. Watts, {\em unpublished}

\end{thebibliography}
\end{document}